**Memristive chaotic circuit for information processing through time**

*Manuel Escudero, Sabina Spiga and Stefano Brivio\**

M. Escudero, S. Spiga, S. Brivio
CNR – IMM, Unit of Agrate Brianza, Agrate Brianza, Italy
E-mail: stefano.brivio@cnr.it

Human brain processes sensory information in real-time with extraordinary efficiency compared to the possibilities of current artificial computing systems. It operates as a complex nonlinear system, composed of interacting dynamic units – neurons and synapses – that processes data-streams as time goes by, i.e. through time, using time as an internal variable. Here we report on a memristor-based compact chaotic circuit included in a computing architecture that can process information through time. We realized a hardware memristive version of the formally simplest chaotic circuit that, thanks to the nonlinearity of the nonvolatile memristor device, evolves with complex dynamics in response to a driving signal. The circuit is used in a single-node reservoir computing scheme to demonstrate nonlinear classification tasks and the processing of data streams through time. These results demonstrate that a simple memristor-based chaotic circuit has the potential to operate as a nonlinear dynamics-based computing system and to process temporal information through time.

# 1. Introduction

Nonlinear dynamics-based processing of information is an alternative to conventional digital computing architectures that are currently posing urgent challenges in terms of energy footprint,[1,2] growing software complexity[3] and personal data protection for computation in the cloud.[4] At an abstract level, computing through nonlinear dynamics is a nature-inspired concept. In general, natural and living systems are able to efficiently cope with different situations thanks to the exploitation of nonlinear dynamic processes.[5] Human brain itself – the ultimate role model for energy efficient computing machine – is considered as a complex network of interacting nonlinear and dynamical entities (neurons, synapses, dendrites, …).

The potential of a nonlinear dynamic computing system stems – as the words say – from its nonlinearity and its internal complex dynamics. Indeed, the nonlinearity in dynamic systems produces complex evolution trajectories, which encompasses many different behaviors, each of which can be selected - without specific programming - for many different kinds of computation.[5] Furthermore, the internal dynamics keeps track of present and – to some extent – past information while computing. Therefore, nonlinear dynamic system intrinsically processes data-streams as time goes by, enabling also efficient real-time operation.[6,7] This is a great advantage compared to digital computers that are constrained to transfer the time variable into a spatial one by storing temporary copies of network states. Such continuous data transfer, known as von Neumann's bottleneck, burdens computational time and efficiency.[8]

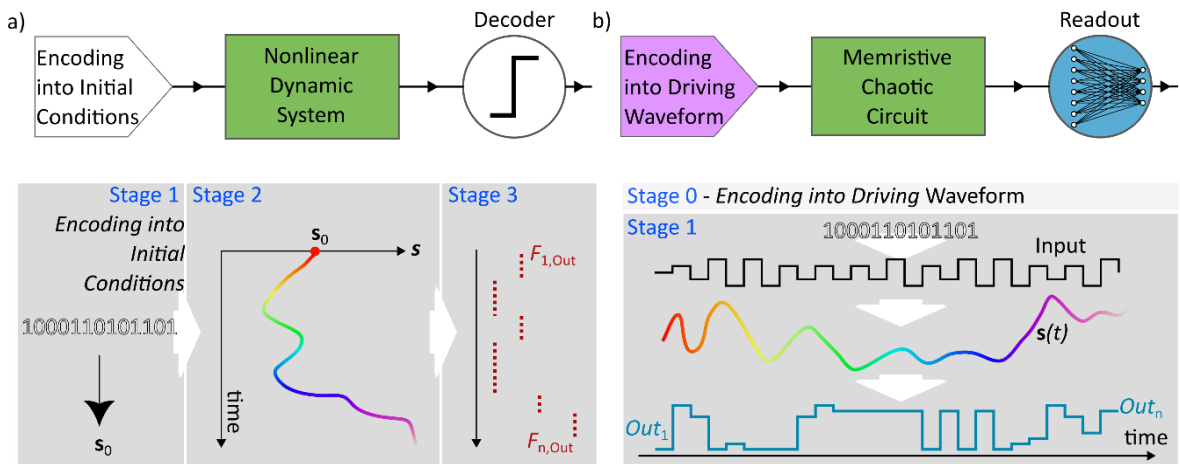

**Figure 1.** Conceptual scheme of two nonlinear dynamic computing systems, one with input encoded in the initial condition of the system (top part (a)) and one - present work - employing a chaotic system with driving signal that encodes the input of the computing task (top part (b)). The computed output is obtained by decoding, in the former case, and through further elaboration, *e.g.* through a classifier or a neural network, in the latter case. In case (a) (bottom part), computing is organized in three stages (encoding, evolution and decoding), where input is encoded as a static signal and the evolution in time of the chaotic system is not associated with an evolution in time of the input signal. Decoding allows obtaining different functions ($F_{i,out}$) at different delays from the start of evolution. In case (b) (bottom part), once the encoding process is defined (stage 0), the input signal can be encoded and fed to the system at time goes by and one task can be computed directly through time in one stage (stage 1) and different functions require different readouts.

Many different paradigms, approaches, proposals and systems - sometimes accompanied by hardware implementations - can be classified as nonlinear dynamic-based computing, like, cellular neural networks,[9] spiking neural networks,[10] oscillator-based computing,[11] self-organized logic gates,[12] self-organized memristive networks[13,14] and many more.[7] In such proposals, the concept of exploiting a complex dynamics leading to chaos or edge of chaos is often addressed. In other cases, chaos is only used as additional factor improving the computation, like for transient chaos computing or chaotic annealing.[15–17]

Recently, the idea of computing through nonlinear dynamic circuits has been developed also with small demonstrators.[18–23] Most of the implementations considers a chaotic circuit for which the initial conditions encode the input data and the processing of the information develops with the time evolution of the trajectories of the state variables of the circuit.[5] This computing scheme is illustrated in Figure 1(a). The computing architecture is composed of an encoder, a nonlinear dynamic system and a decoder. Computing is performed in three stages: the initialization of the system into static initial conditions that encode the input signal (stage 1), the evolution in time of the nonlinear system (stage 2) and the decoding of the output (stage 3). This scheme allows the implementation of implementing different functions for different evolution times of the chaotic system. On the downside, it does not take full advantage of the dynamic memory of the system, *i.e.* the ability to process information through time. In the system architecture of Figure 1a, the evolution in time of the chaotic system is somewhat transversal to a possible time evolution of the input signal (hence the representation in Figure 1a), because each input signal is coded in the initial condition of the system.

In this work, we use a chaotic circuit to demonstrate the computation through time of a nonlinear dynamic system. The conceptual computing scheme is reported in Figure 1b. Three blocks are needed: an encoding block, a chaotic system and a readout, that, in our case, is a classification or regression layer and must be trained. The system operation is sketched in the bottom part of Figure 1b. Once defined a suitable encoding procedure and system initialization, the input is fed to the chaotic system as a driving signal in real-time and elaborated by the readout. The architecture implements a computation through time, using the time evolution of the chaotic system as a sort of working memory and formally collapsing the computation in only one stage. Therefore, the computing system of Figure 1b works in a completely different way from the counterpart in Figure 1a, despite the similar organization of the conceptual blocks.

The proposed architecture of Figure 1b can be considered as a single-node implementation of a reservoir computing system as it is composed by a single dynamical entity and a readout layer that is the only part subjected to training.[24] Usual implementation of single node reservoirs, *e.g.* in photonics[25,26] or electronics[22,27], make use of delays that are not easily implemented in electronic circuits and make the system hard to control.[28] Furthermore, in the architecture proposed in this work, the dynamic system is composed of a memristor-based version of the (formally) simplest chaotic circuit that is *non-autonomous*,[29] *i.e.* it requires a driving signal that can be used for feeding the time-varying input to the circuit. The nonlinearity of the circuit, which determines the chaotic evolution, is provided by the tunable nonlinear current-voltage characteristic (*i.e.* its voltage-dependent resistance) of a memristive device.

In the present work, we realize the circuit in hardware and prove its chaotic operation. We demonstrate the two basic ingredients of the proposed computing scheme: *(i)* nonlinear classification and *(ii)* computation through time, considering the implementation of Boolean functions, which are usual test benches for innovative computing systems.[30–32]

Contextually, we take advantage of the nonvolatile memristor as a knob to optimize the computing performance.

## 2. Results

### 2.1 Non-Autonomous Memristive Chaotic Circuit with Periodic Driving Signals

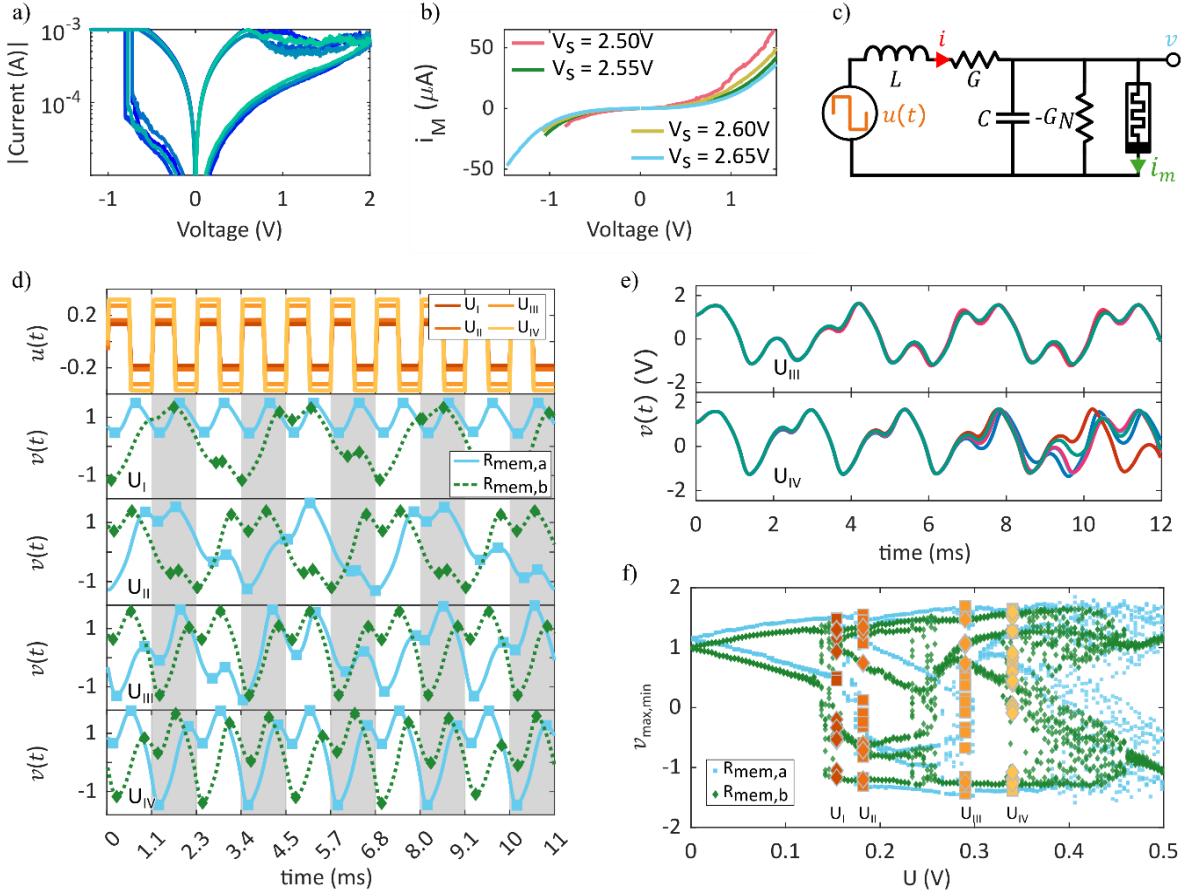

**Figure 2.** A nonvolatile Pt(top electrode)/HfO$_2$/TiN(bottom electrode) memristive device can be programmed through voltage stimuli to low and high resistance states (a). Several stable resistance states that have different current-voltage nonlinear characteristics (b) can be programmed by tuning the stop voltage (V$_S$) of the positive sweep. (c) The device is included in the chaotic circuit that is driven by periodic driving signals, *e.g.* a square wave. (d) Experimental response of the circuit to a square wave with increasing amplitudes (U$_I$ = 0.161 V, U$_{II}$ = 0.188 V, U$_{III}$ = 0.299 V, U$_{IV}$ = 0.346 V). Continuous blue and dashed green lines correspond to different representative memristive states, R$_{mem,a}$ = 968 kΩ and R$_{mem,b}$ = 747 kΩ, respectively. (e) Examples of divergence of the chaotic output of the circuit in several repetitions with same initial condition (within the experimental accuracy), same memristor state, 465 kΩ and different amplitudes, U$_{III}$ and U$_{IV}$ in top and bottom panels, respectively. (f) The local maxima and minima of the circuit output voltage (as indicated in (d) with blue squares and green rhombs for R$_{mem,a}$ and R$_{mem,b}$) are collected as a function of the voltage amplitude of the driving square wave to draw a bifurcation plot of the circuit and for two different memristive device states (the same as in (d)). Local maxima and minima corresponding to driving signals with amplitudes U$_I$, U$_{II}$, U$_{III}$ and U$_{IV}$ in (d) are reported in colors from red to orange and to yellow as in the top plot of panel (d) and as squares and rhombs for the R$_{mem,a}$ and R$_{mem,b}$. An additional bifurcation diagram obtained with the same memristor state as in (e) is reported in the Supplementary Figure S1.

In the present work, we take advantage of the nonlinear current-voltage characteristic of nonvolatile Pt/HfO$_2$/TiN memristive devices (memristors for the sake of brevity) able to switch the resistance across their two terminals. Figure 2a shows the response of the device to bipolar quasi static voltage ramps leading to transitions from high to low resistance states (and vice versa) for negative (positive) voltage polarity. The device can be set into several stable states with different nonlinearity of their current-voltage characteristics (Figure 2b) by controlling the stop voltage ($V_s$) of the positive sweep or the current compliance in the negative sweep.[33] We choose to indicate a memristor state with its resistance value read at a low voltage of 100 mV. Figure 2c shows the electric scheme of the circuit exploiting the nonlinearity of the memristive devices, which is derived from the (formally) simplest chaotic circuit as proposed by Murali *et al*.[29] The circuit is described by the following equations

$$C\frac{dv(t)}{dt} = i(t) - G_N - i_m(v), \qquad (1)$$

$$L\frac{di(t)}{dt} = u(t) - i(t) \cdot R - v(t), \qquad (2)$$

with $u(t)$ the input waveform, $v(t)$ the circuit output, coinciding with the voltage dropping on the memristor, and $i$ is the current flowing through the inductor. $i(t)$ and $v(t)$ are the state variables of the circuit. The other quantities are the circuit components defined in Figure 2c. The unique nonlinearity in the circuit is that of the memristor, appearing eq. (1) through the current-voltage characteristic $i_m(v)$. Such current-voltage relationship can be fit through polynomial equation to accurately simulate the circuit response as already discussed in ref. [34,35], where the same authors also discuss the design rules and the considerations about the use of a memristive device in the autonomous (*i.e.*, not requiring any driving signal) version of the same chaotic circuit. Such design rules apply similarly to the non-autonomous circuit (*i.e.*, requiring an driving signal) employed in the present work and reported in Figure 2c.[34,35]

When fed with a periodic driving signal as the square wave in Figure 2d (top panel), the non-autonomous circuit oscillates with complex dynamics. The circuit output varies as a function of the amplitude of the driving signal as shown by the bottom panels in Figure 2d. The delivery of the waveform is preceded by a voltage pulse that initializes the circuit always to the same equilibrium state at $v > 0$, as described in the Experimental Section.

The circuit output can be tuned by changing the memristor state as shown by the results in continuous (blue) and dashed (green) lines, corresponding to different representative low-voltage resistance values ($R_{mem,a}$ and $R_{mem,b}$). From a qualitative point of view, one can appreciate periodic oscillations (for U = U$_I$) with the same period as the driving waveform for $R_{mem,a}$. With the different state $R_{mem,b}$, the circuit output is periodic with a period that is tripled compared to the driving signal. Increasing the driving voltage amplitude (from U$_I$ to U$_{II}$, blue lines), also for $R_{mem,a}$, complex oscillations occur at the output with period multiplication. A further amplitude increase (from U$_{II}$ to U$_{III}$, blue lines, $R_{mem,a}$) results in a nonperiodic output. When the measurements reported in Figure 2d are repeated several times, the chaotic nature of the circuit behavior becomes evident. Figure 2e shows an ensemble of several repetitions of the output waveforms for U = U$_{III}$ and U = U$_{IV}$. The traces follow the same trajectory for a certain time after the start of the recording (about 6 ms) and then they diverge towards different paths. This fact indicates that even if the circuit is in a chaotic regime the output trajectories are repeatable for a certain limited time. The recording of data as in Figure 2e for different values of the voltage amplitude of the driving waveform allows building a bifurcation diagram that describes the overall behavior of the circuit, as the one reported in Figure 2f. The plot collects the critical points – local minima and maxima ($v_{min,max}$) – of experimental output waveforms as a function of the amplitude of the driving waveform (local maxima and minima are also indicated in panel (d) with squares and rhombs for $R_{mem,a}$ and $R_{mem,b}$, respectively). For low amplitudes of the driving signal (U ≲ 0.13 V), the circuit state variable $v$ periodically oscillates around the equilibrium point at $v > 0$ defined by the initialization. Therefore, always the same values (within the experimental accuracy) for the local maximum and minimum are reached during the oscillations, leading to only two points for a fixed driving amplitude in the bifurcation plot for U ≲ 0.13 V. For high amplitude values (U ≳ 0.13 V), bifurcations occur and the oscillations become complex – involving orbits around both equilibrium points at opposite polarities. Oscillations can be periodic with period multiplication, as indicated by the presence of multiple points for fixed driving amplitude (multiple local minima and maxima of $v$) and they can be nonperiodic and chaotic, as indicated by clouds of points. Green and blue symbols correspond to the different memristor states $R_{mem,a}$ and $R_{mem,b}$, indicating that the circuit response can be modified in a nonvolatile manner through proper device programming. The local critical points of the output waveforms obtained with driving signal amplitudes $U_I$-$U_{IV}$ are indicated by large squares and rhombs for $R_{mem,a}$ and $R_{mem,b}$ with color ranging from red to yellow.

## 2.2 Nonlinear classification tasks in a reservoir-like computation

The memristive chaotic circuit is used as a computing engine in a reservoir computing scheme[36,37] as represented in Figure 3a. The system is tested against linear and nonlinear classification tasks that are the implementation of Boolean functions with 2 and 3 input bits. As only one signal drives the circuit operation (as usual in single-node reservoir systems[26,27]), the multibit Boolean input must be encoded into one variable. Figure 3b reports the employed encoding procedure for the 2-input-bit case which linearly maps the bit sequence into the voltage amplitude of the driving square wave.

The system operates as follows (please refer to Figure 3a). A square wave with an amplitude encoding a specific bit sequence (00, 01, 10 or 11 in case of 2-input-bit function) drives the circuit after the initialization procedure described in the Experimental Section. The circuit evolves and its output trace, $v(t)$, is sampled (with an oscilloscope in our experiments), and a dataset is therefore collected for different input voltage and memristor state (full data are reported in the Supplementary Figure S2). The samples, $v(t_i)$, are used as the input array of a linear classifier, *i.e.* the readout layer of the reservoir architecture, that is trained by a linear support vector machine algorithm. The testing (inference) is performed driving again the circuit and through software implementation of the trained readout.

The encoding procedure described above can be extended to an arbitrary number of input bits and it corresponds to a sort of compression of the input, which may be $n$-dimensional for the generic case of $n$ input bits. It must be noticed that some Boolean functions that correspond to a linear classification task in their original $n$ dimensional space may be translated into a nonlinear problem when encoded into a 1-dimensional space. This fact is described in Figure 3c, which shows the input/output space for 4 representative Boolean logic functions ("*a* AND *b*", "NOT(*a*) AND *b*", "*a* AND NOT(*b*)" and the exclusive disjunction between *a* and *b*, *i.e.* "*a* XOR *b*"), where the input bits are ordered horizontally, in agreement with the encoding process of Figure 3b, and circles (○) or crosses (×) represent 1 and 0 expected outputs. The only linearly separable function is the "AND" function, because a single line (dashed line in the figure) can distinguish between expected output 1 and 0. The remaining reported functions require at least two lines indicating they corresponds to a nonlinear classification task. It can be graphically verified that the functions "NOT(*a*) AND *b*" and "*a* AND NOT(*b*)" correspond to a linear classification problem in the [*a*,*b*] 2-dimensional space, while XOR function is a nonlinear function both before and after encoding.

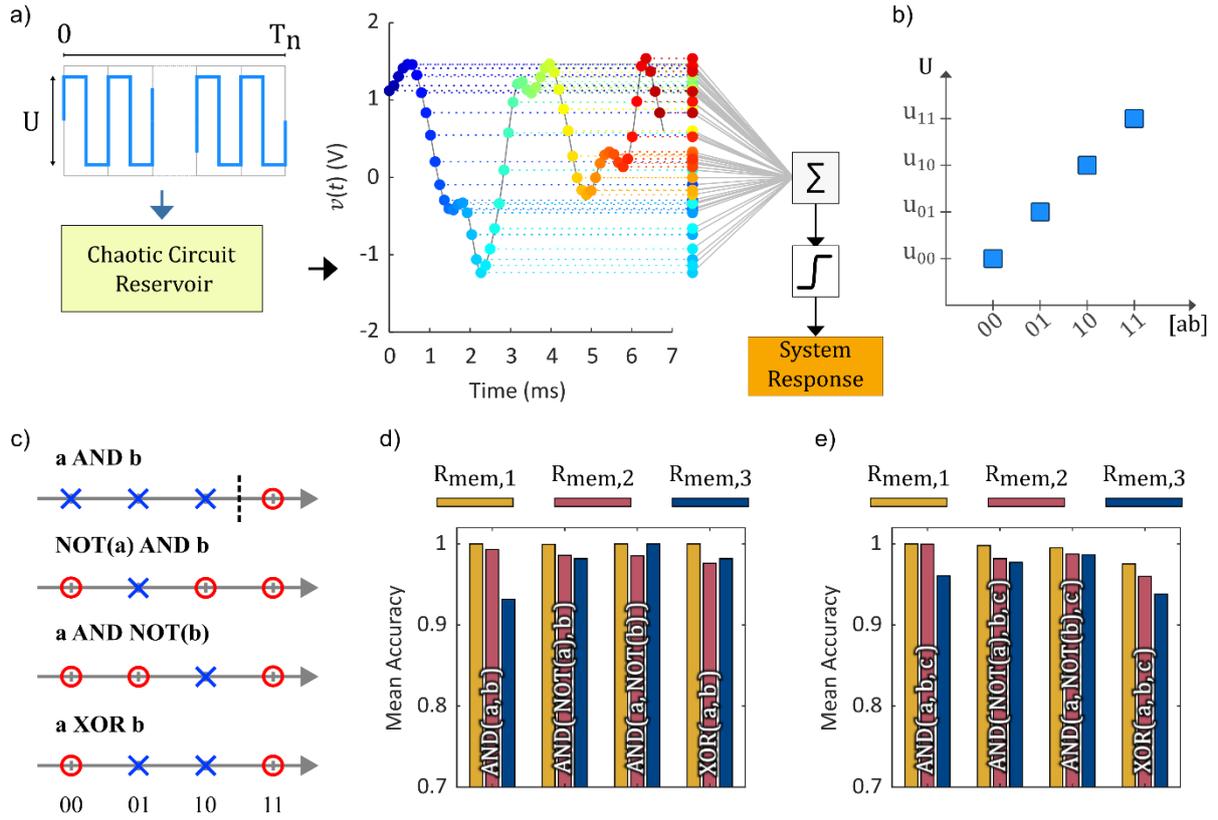

**Figure 3.** (a) Scheme of the computing system that includes (from left to right) the driving signal fed to the chaotic circuit, the sampling of its output, $v(t)$, versus time (colored filled symbols), that is eventually delivered to a linear classifier that operates a weighted sum and thresholding operation and returns the system response. (b) Example of simple masking strategy that encodes bit sequences into the amplitude of a periodic signal. The bit sequences, [ab], represent the input of a 2-dimensional logic function to be implemented by the system in (a). In the case of a 2-dimensional logic function, the 4 possible bit sequences are 00, 01, 10, 11 that are coded into 4 different amplitudes of the periodic driving signal. (c) Explanation of linear and nonlinear separability for four 2-input functions where the input is compressed into one variable according to the encoding reported in (b). "X" and "O" correspond to expected output "1" and "0", respectively. "AND" function is linearly separable within the 1D space (the dashed line represents a possible separation threshold), functions "NOT(a) AND b", "a AND NOT(b)" and "a XOR b" are separable only through a nonlinear expansion into a higher dimensional space. (c-d) Mean accuracy of four logic functions with 2 and 3 input bits, respectively, for three different values of the memristor state ($R_{mem,1}$ = 465 k$\Omega$, $R_{mem,2}$ = 755 k$\Omega$ and $R_{mem,3}$ = 1.04 M$\Omega$). Additional 2- and 3-input-bit functions are reported in Supplementary Figure S3. Computation is performed considering all the samples acquired by the oscilloscope (1000 points) but they can be significantly reduced through a pruning procedure without accuracy degradation as described in Supplementary Figure S4.

In the proposed computing architecture, the reservoir projects the input into a higher dimensional space so that the Boolean functions – also those corresponding to a nonlinear classification – can be correctly implemented by the linear readout layer. Figure 3d-e reports the values of classification accuracy for selected Boolean functions with 2 ([$a,b$]) and 3 ([$a,b,c$]) inputs, respectively. For each function and number of inputs, the experiments are performed with 3 different memristor states ($R_{mem,1-3}$). The accuracy values are high and close to perfection, at least for the 2-input-bits case, for the memristor state $R_{mem,1}$. The different values of the accuracy for different memristor states indicate that the memristor can be used as a knob to optimize the system performance. The classification is performed considering all the sampling points acquired by the oscilloscope used in the experiments (1000 points). However, consistent pruning towards a few tens of points is possible without any accuracy degradation as illustrated by Supplementary Figure S4. The inference step for the XOR Boolean function is also implemented in hardware through a multiply and accumulate operation performed by an array of Pt/HfO$_2$/TiN devices whose conductance values map the weight optimized by software training. The hardware inference reaches an accuracy of 95% as illustrated in Supplementary Figure S5.

It must be mentioned that the results have been obtained without any special optimization of the system and encoding procedure. The values of the circuit components have been chosen according to the design rules explained in ref. [34,35] and summarized in the Experimental Section.

**2.3 Reservoir-like computing through time**

The examples discussed in the previous section demonstrate the ability of the circuit to perform nonlinear classification tasks. Figure 4 demonstrates an example of information processing through time, that is the most peculiar feature of the reservoir computing concept and of the computing through nonlinear dynamic system in general. The operation of the system is described in Figure 4a. In this case, input bits are provided in a time sequence where high and low bits are coded in the high and low amplitude values of one period of a square wave, respectively. The system is meant to implement *n*-input-bit Boolean functions (*n* = 3 in the figure indicated as a *sliding window* of bit processing) while input bits are presented through time to the system. The waveform encoding the input bit stream is fed to the circuit and the output is sampled and sent to the readout layer that is trained through the ridge regression method to return the response of specific Boolean functions (XOR in the example). The response of a Boolean function operating on generic input bits in periods *i*-2, *i*-1 and *i* is returned *i*-th period of the output signal of the overall system. Figure 4b illustrates an example of bit stream, X($t$) (top panel), its encoding into an amplitude modulated square wave, $u(t)$ (middle panel), and the resulting reservoir output ($v(t)$, bottom panel). A careful inspection of the bottom panel of Figure 4b allows noting that the reservoir output does not depend only on the instantaneous input currently fed to it, but it is sensitive also to the past input waveform. To illustrate this fact two sequences of input bits are highlighted by yellowish rectangles (bit sequences 101 and 001 in periods 9-11 and 14-16). They start with a different bit value (periods 9 and 14) and go on with two identical bit values (periods 10-11 and 15-16). In response to the two 3-bit sequences, the reservoir output follows different trajectories up to the corresponding third periods, despite the only difference is in the first period. Therefore, the reservoir has a fading memory that temporarily keeps track of past information. Figure 4c reports (filled black circles) the response expected from the same input bit sequence as in Figure 4b (reported also on top of Figure 4c) and the computed system response (empty red squares) for 2-input-bit functions (OR, AND and XOR) and XOR functions with 3 and 4 input-bits. The overlapping of the markers indicates correct function evaluation. The system allows reproducing the reported representative Boolean functions with good accuracy. The same computation is performed for several bit streams and the resulting average accuracy is reported in Figure 4d for functions OR, AND and XOR with 2,3 and 4 inputs (different colors) as a function of the memristor state. First, it can be noticed that the 4-input-bit functions are generally associated with a lower accuracy (more evident for OR and XOR functions) throughout the whole investigated range of the memristor state. Conversely, for 2- and 3-input-bit functions, the accuracy of the system can be optimized by a careful choice of the memristor state. For instance, for a low voltage resistance around 500 kΩ the Boolean functions show optimum accuracy.

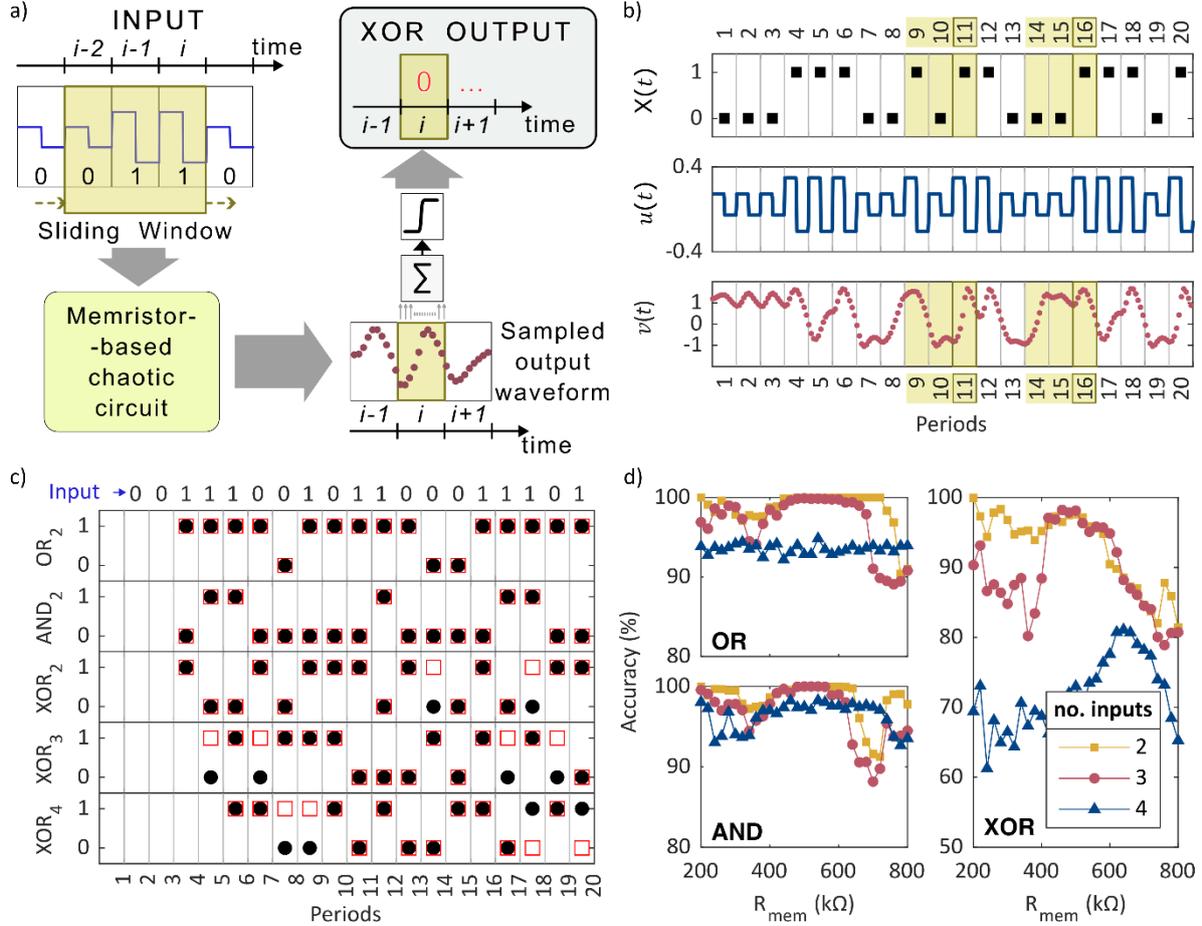

**Figure 4.** (a) Scheme of the computing system implementing sequential Boolean function over bit streams. Low and high bit values are implemented in low and high amplitudes of one period (T = 1.134 s) of a square waveform ($U_{LOW}$ = 0.1V, $U_{HIGH}$ = 0.25 V). An offset of 50 mV is set to select a preferential equilibrium point in the reservoir output trajectory as discussed in the Experimental Section. The input waveform is fed to the memristor-based chaotic circuit. Its output is acquired, sampled and delivered to the readout layer that computes a Boolean function, in the specific example, a XOR function on 3 sequential bits. The function response to the input bit fed in periods [$i$-2,$i$] is returned during period $i$, as indicated by the yellowish rectangles. As the bits are fed through time, the system computes a Boolean function over a sort of sliding window. (b) Input bit stream, $X(t)$ (top panel), encoding driving waveform, $u(t)$ (middle panel), and experimental circuit output waveform, $v(t)$ (bottom panel) as a function of time. Yellowish rectangles indicate two 3-bit sequences, 101 and 001 in periods 9-11 and 14-16 that start with a different bit value and go on with the same bit values. The corresponding output of the circuit is still different in the third periods (compare $v(t)$ in periods 11 and 16), indicating that the circuit holds the memory of a different input for a time as long as, at least, three periods. (c) Expected system response (filled black circles) and classification (empty red squares) computed by the system for different Boolean functions: 2-input OR, AND and XOR (OR2, AND2 and XOR2); 3-input XOR (XOR3) and 4-input XOR (XOR4). The input bit sequence is reported on top. The classification is performed as soon as enough bits are acquired depending on the number of inputs required by a specific function. Overlapping markers indicate a correct function evaluation. (d) Classification accuracy for OR, AND and XOR for 2 to 4 inputs as a function of the memristor resistance ($R_{mem}$).

## 3. Discussion and Outlook

The present work demonstrates the use of a memristor-based computing circuit for the implementation of nonlinear classification tasks through time. A single-node reservoir architecture is used as a straightforward example of computation through a nonlinear dynamical system. The complex dynamics of the circuit, resulting from the interaction between nonlinear (the memristor) and dynamical elements (capacitor and inductor), provides the ability of nonlinear classification processing. The state variables of the circuit, given by voltage and current variables of dynamic components, evolve in time in response to the circuit driving signal and to their dynamic reciprocal interaction. For this reason, the internal circuit dynamics operates as a sort of temporary working memory that is necessary for the processing of information through time. The memristor device included in the circuit can be programmed in specific nonvolatile state to optimize the system performance. In the next subsections, specific key points of the work are discussed.

### 3.1. Input Encoding

In general, for any machine learning problem and specifically in single-node reservoir architectures, input encoding is critical for accurate results in challenging computing tasks.[24,25] A very straightforward encoding technique is intentionally adopted in the present work to demonstrate the potential of the computing concept based on a chaotic circuit. For the demonstration of nonlinear classification, the encoding produces a data compression that makes some linear Boolean function nonlinear (see Figure 3c). No attempt to recover such a drawback, *e.g.* including some nonlinearity in the encoding strategy, is put in place in the present work which further demonstrates the processing ability of the system. The only considered heuristic optimization of the encoding technique is to place the voltage range of the driving signal across the region where the bifurcation diagram (Figure 2f) is complex, in order to favor the nonlinear expansion of the input signal. Therefore, also concerning the input encoding, huge optimization work is possible to improve computing performance.

### 3.2 Repeatability

The use of a chaotic circuit to perform computation is moved by the intuition that complex dynamics and system close to the edge-of-chaos expand the information included in the input and support further elaboration.[5,38,39] On the other hand, chaotic trajectories are not repeatable for an indefinite time interval as demonstrated experimentally with reference to Figure 2e, which is unwanted in an information processing system. A compromise between complexity and repeatability must be found in order to improve the computing capacity. There are several possible knobs that can be used for this aim. Figure 2e demonstrates that circuit trajectories become unrepeatable only after a certain delay. For this reason, the training process itself implicitly selects the most reliable samples in the circuit output trace. Indeed, it associates large weights with the most significant samples and low weights with those values that are not significant because highly variable and not representative of a specific behavior. Furthermore, the memristor nonlinearity is another possible variable that can be used to choose the optimum computing performances. Indeed, Figure 3d-e and Figure 4d show that optimum memristor states exist for maximum classification accuracy. However, also in this case, a large room for improvement is available which deserves a combined theoretical/experimental effort.

### 3.3 Circuit Design

The nonlinear dynamic system presented in this work is the simplest non-autonomous chaotic circuit modified with the inclusion of a memristive device, which, by itself, has been the subject of intense studies.[40–43] With previous works by the same authors[34,35] and the present one, memristive versions of both autonomous and non-autonomous chaotic circuits are experimentally demonstrated. The memristor device provides the nonlinearity that is fundamental for the insurgence of chaos, together with the possibility of tuning it with a nonvolatile retention.[44,45] The nonlinear current-voltage characteristics of the filamentary nonvolatile Pt/HfO$_2$/TiN device[44–46] is ascribed to an electronic conduction mechanism that occurs in part through field- and temperature-activated conduction through networks of defects (*e.g.* Poole-Frenkel,[44,47] trap-assisted tunnelling[48–50]) and in part because of the presence of barriers at the metal/oxide interfaces.[51] To the best of authors' knowledge, the nonlinear current-voltage characteristic of memristive devices has never been engineered before in view of their exploitation in chaotic circuits. However, some considerations can be drawn which are useful for future optimization. Indeed, devices with interface-dominated conduction, often associated with area-dependent switching, usually display larger dependence of resistance on the applied voltage and nonlinearity, which simplifies the circuit design. Area-dependent switching devices, compared to filamentary ones, also usually feature (*i*) higher switching voltages,[8,52] which allow the use of a wider range of driving voltages in the circuit,[35] and (*ii*) lower cycle-to-cycle variability and noise,[8,52] which improves computing accuracy. On the downside, area-dependent devices exhibit poorer retention than filamentary ones,[8,52] which hinders a long term circuit operation after device programming.

The choice of the (formally) simplest chaotic circuit serves the scope of demonstrating the potential of the computational scheme, but there is room for accuracy improvement implementing novel circuits designed through the guidance of nonlinear dynamics theory. As a matter of fact, other dynamic circuits have been proposed for computing especially following the scheme of input encoding into the initial condition of the circuit state variable (see Figure 1a).[18,21] Volatile memristive devices with insulator-to-metal transition have been demonstrated to show chaotic dynamics. Also in this case, they have been proposed for non-temporal tasks, like optimization/minimization problems through chaotic annealing.[17]

In summary, we enrich the formally simplest non-autonomous chaotic circuit with the tunable and nonvolatile nonlinearity of a memristive device to realize a nonlinear dynamic computing system able to process temporal information through time exploiting the circuit internal variable as short-term working memory.

## 4. Experimental Section

### 4.1 Device preparation and testing

50 nm Pt/5.5 nm HfO$_2$/40 nm TiN memristive devices are prepared by sputtering deposition of metal films and atomic layer deposition of the oxide and optical lithography, as discussed in details in ref. [44]. The size of the device is 40x40 µm$^2$. A forming sweep at negative voltage polarity applied to the top Pt electrode is needed to initiate the switching. Switching operation can be prolonged for at least thousands of cycles is quasi-static conditions.[46] A retention time of 10 years is demonstrated at 170°C.[44]

### 4.2 Circuit Design and Realization

Our circuit is designed to exhibit similar dynamics to the original circuit[29], *i.e.*, various periodic and chaotic orbits. More specifically, we guarantee the same stability conditions in the system of eq. (1)-(2) by imposing two asymptotically stable points and a saddle point in the autonomous case, *i.e.*, $u(t) = 0$. Assuming a previously developed model, $i_m(t)$, for the memristor current-voltage characteristic[33] across the range of low voltage resistances [0.1, 1] MΩ, the resulting equation to size the circuit impedances are

$$R = \frac{1}{k} \frac{\partial i_m}{\partial v}\bigg|_{v=0, R_{mem,min}}, \quad (3)$$

$$R_N = R/(1 + k), \qquad (4)$$

$$L = CR^2. \qquad (5)$$

$R$ depends on the maximum linear conductance of the memristor at given range, and $k$ (in this work, $k = 5$) is a term that helps to adjust the voltage levels applied to our device and to allow some tolerance margin for the memristor state variability. The circuit is therefore designed with the following values $C$ =10 pF, $R$ =13.54 kΩ, $L$ =1.833 H and $R_N$ = 11.28 kΩ.

The circuit is implemented in a breadboard and connected to the Pt/ HfO$_2$/TiN device terminals through a probe station. The inductor is emulated by a gyrator circuit, allowing to implement a high inductance value, while the negative conductance G$_N$ is provided by a simple operational amplifier configuration.

### 4.3 System Measurements

The memristive devices are programmed and measured using a Keysight B1500A semiconductor parameter analyzer. The low voltage resistance of the devices is specifically obtained at 100 mV after the programming procedure.

System measurements with the physical circuit are performed along with an arbitrary waveform generator 33220A and an oscilloscope MSO6430. The waveform generator is used for circuit control and generation of the circuit driving signal, while both the circuit driving signal and output are acquired through the oscilloscope. The measurements are automated with LabVIEW, which is in charge of encoding the input data into the driving waveform, synchronizing it with the circuit output measurement, and setting the sampling frequency.

The generated waveform comprises three sequential steps: triggering pulse, initialization pulse, and input data signal. The triggering pulse forces the synchronization between the waveform generator and the oscilloscope to maximize the oscilloscope acquisition window, with a 0.5 ms wide pulse and an amplitude 0.2 V higher than the amplitude of the input data signal amplitude. The initialization pulse sets the circuit initial conditions. By applying a 0.5 ms wide pulse of 0.2 (-0.2) V amplitude, the output starts at a fixed positive (negative) voltage value. Finally, the waveform generator produces the square signal that encodes the input data, consisting of multiple 1.134 ms-long periods (20 or more) and variable amplitude according to the input data. It is worth noticing that, in the case of the time-varying signal evaluation, an offset of 10 mV was added to the input data signal segment, to force the circuit to only have a single stable equilibrium point and thus granting fading memory to the chaotic circuit.

### 4.4 Software Training and Inference

The training and inference of the readout layer are carried out with a MATLAB script. In the case of the processing of periodic inputs (Figure 3), a linear support vector machine classifier was used for training the readout. When processing time-varying inputs (Figure 4), the built-in ridge regressor was used instead, with ridge (or regularization) parameter of 10$^{-3}$. While both methods reach similar accuracies, the support vector machine was superior as it better dealt with unreliable output samples due to noise and/or chaos.  For the computation with periodic input, the dataset is divided into 80% for training and 20% for validation, while for the computation of time-varying signals, the dataset is equally split for training and validation.

**Acknowledgements**

The work is partially funded by the PRIN2017-MIUR project Analogue Computing with Dynamic Switching Memristor Oscillators: Theory, Devices and Applications (COSMO, Prot.

2017LSCR4K) and partially funded by Ministero delle Imprese e del Made in Italy (MIMIT) under IPCEI Microelettronica 2, project MicroTech_for_Green.

# Supporting Information

**Memristive chaotic circuit for information processing through time**

*Manuel Escudero, Sabina Spiga and Stefano Brivio\**

**S1. Bifurcation Diagram**

The bifurcation diagram of the circuit with a memristor device programmed in state with R = 465 kΩ is reported in Figure S1.

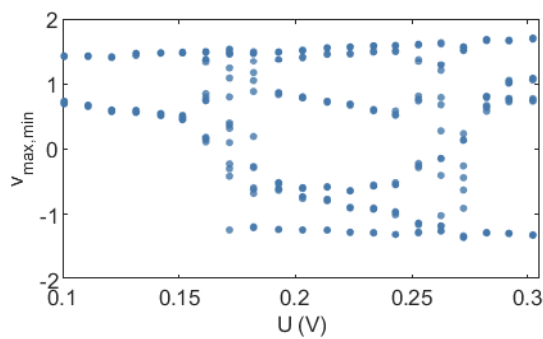

**Figure S1.** Bifurcation diagram of the circuit for the memristor state 465 kΩ

**S2. Nonlinear classification dataset**

The dataset used for the implementation of the Boolean functions reported in Figure 3. For illustration purposes, we report in Figure S2 the subset of the dataset employed for learning 2-input Boolean functions. The measurements illustrate the periodic and non-periodic responses, as well as the divergence rate between repeated evaluations in the case of non-periodic responses.

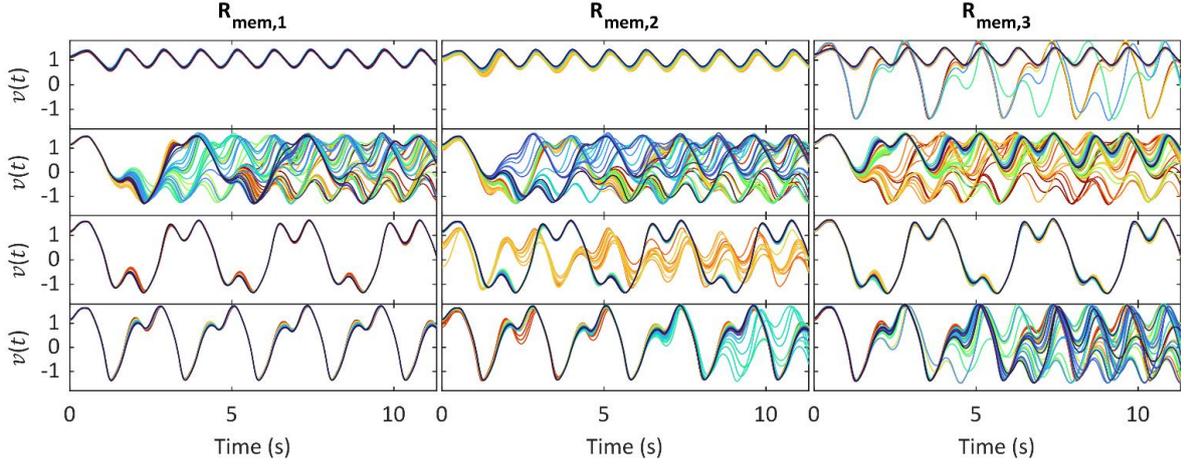

**Figure S2.** Set of 50 circuit output traces (different colors), $v(t)$, as a function of time in response to square waveforms with different amplitudes ($U_I$ = 0.161 V, $U_{II}$ = 0.188 V, $U_{III}$ = 0.299 V, $U_{IV}$ = 0.346 V, different panel rows, corresponding to different bit combinations 00, 01, 10, 11) and for different memristor states ($R_{mem,1}$ = 465 kΩ, $R_{mem,2}$ = 755 kΩ and $R_{mem,3}$ = 1.04 MΩ, different panel columns).

### S3. Implementation of additional Boolean functions

We extend the results of Figure 3 of the manuscript by reporting here the accuracy of additional 3-input Boolean functions. Only the majority function (MAJ) is linearly separable in the 3-input Boolean space, while the multiplexer (MUX), the XORAND and ANDXOR function are nonlinearly separable. As in Figure 3, the results confirm results close to perfect accuracy with the memristor state $R_{mem,1}$.

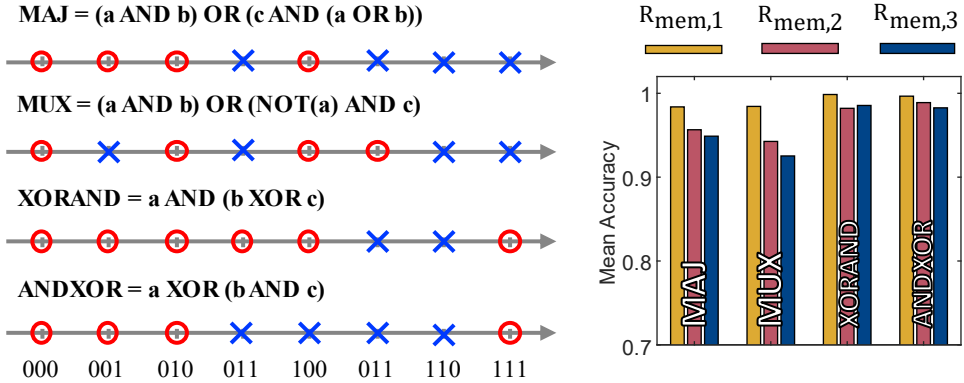

**Figure S3.** Mean accuracy of logic functions with 3 input bits using periodic inputs, for three different values of the memristor state ($R_{mem,1}$ = 465 kΩ, $R_{mem,2}$ = 755 kΩ and $R_{mem,3}$ = 1.04 MΩ).

### S4. Weight pruning and hardware implementation of readout layer

The results reported in Figure 3 of the main manuscript are obtained through software implementation of the readout layer and all the samples acquired by the oscilloscope (1000 points). Deep pruning of the readout layer allows to reduce the number of processing points to less than few tens of points without sacrificing the computing accuracy. Figure S4 shows the accuracy as a function of the number of pruned weights (bottom *x* axis) – or dually the number of remaining weights (top *x* axis) – for 2- and 3-input-bit Boolean functions (panel (a) and (b), respectively, for the three memristor states $R_{mem,1-3}$.

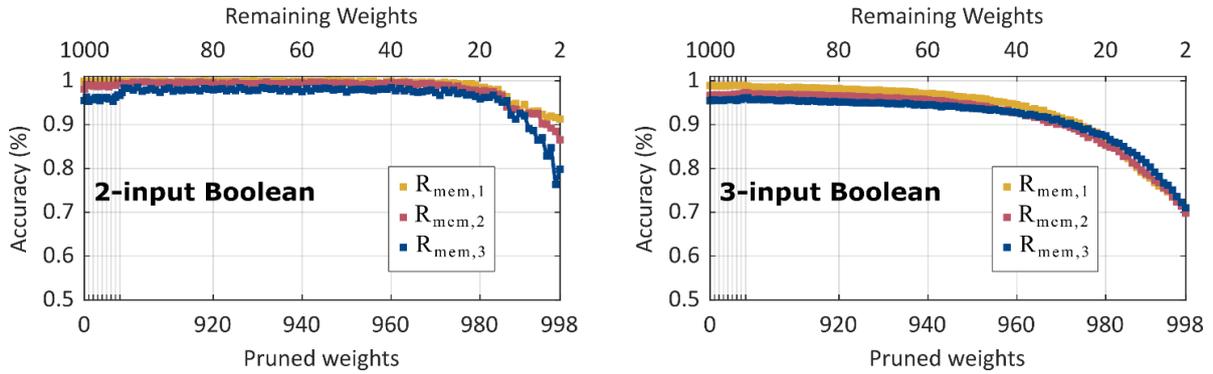

**Figure S4.** Classification accuracy averaged over many 2-input-bit and 3-input-bit Boolean functions (panel (a) and (b), respectively) as a function of the number of pruned weights (bottom *x* axis) or dually as a function of the remaining weights (top *x* axis) for different memristor states ($R_{mem,1}$, $R_{mem,2}$, $R_{mem,3}$, different panel columns).

The pruning process allows reducing the number of weights down to 4 with a mild degradation of the performance for 2-input-bit functions (Figure S4a) especially for $R_{mem,1}$. In this special case, the readout layer multiply and accumulate (MAC) operation is implemented in hardware for the inference stage, as a proof of concept. The crosspoint architecture employed for such in-memory computation is reported in Figure S5a. The devices used are the same as the one included in the chaotic circuit. The conductance values of the memristors weights the current flowing through them and the overall current measured by the parallel of 4 memristive devices is the sum of each current. Therefore, the architecture implements a MAC operation.[1–4] Through the program-and-verify technique, illustrated in Figure S5b, their conductance is set to values that are proportional to the weight resulting from the readout training performed on a computer. The program-and-verify algorithm is based on the iterative increase of the compliance current used during the SET operation similarly to other works present in the literature.[5] The algorithm stops when the conductance value reaches a value higher than an identified threshold (a value 5% lower than the target conductance is used in the experiment).

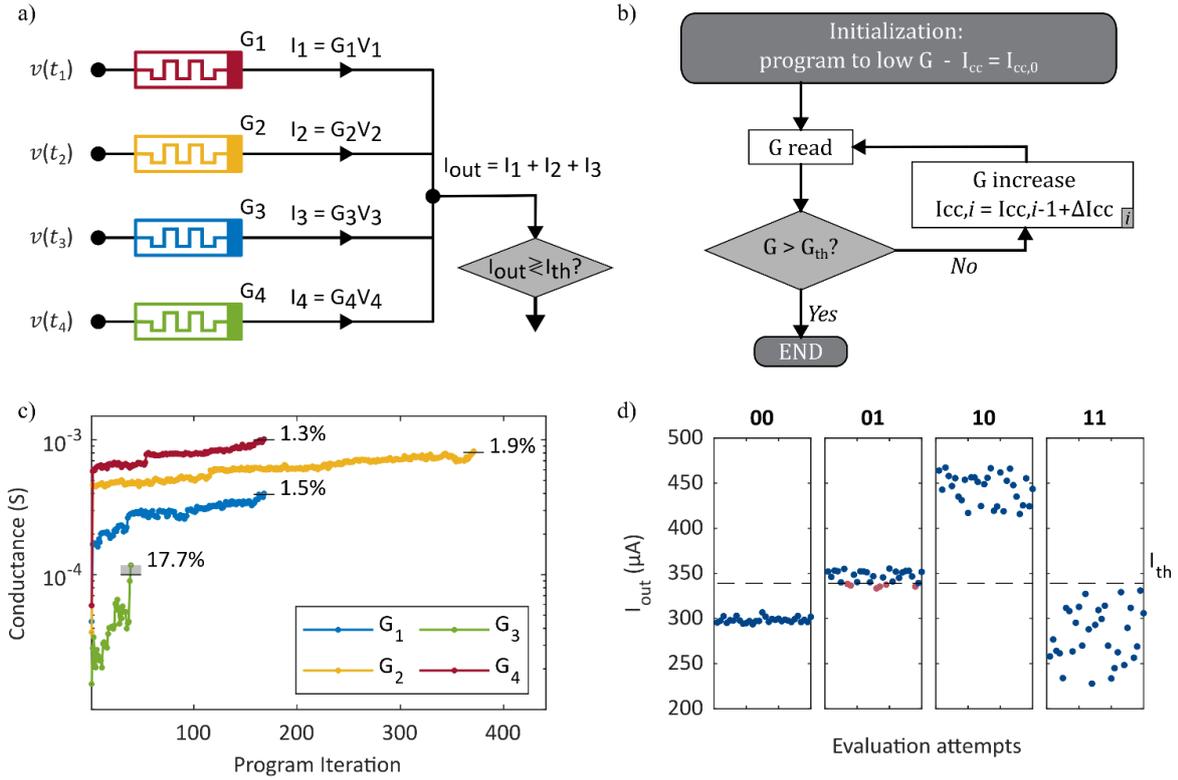

**Figure S5.** (a) Memristive crosspoint architecture implementing the scalar product between the 4 voltage values of the trace of the reservoir output, $v(t_n)$, selected by pruning, and the conductance values of 4 memristive devices, $G_1$-$G_4$. The resulting current, $I_{tot}$, is then compared to a certain threshold to assess the network response. (b) Program-and-verify scheme for the accurate programming of memristor conductance values. Devices are first initialized to a low conductance value through the application of a low compliance current, $I_{cc,0}$, during a SET operation. The compliance value is iteratively increased by a constant amount, $\Delta_{Icc}$, until the conductance read at low voltage (50 mV) becomes higher than the conductance threshold, $G_{th}$, that is fixed to 5% lower than the target value. (c) Conductance as a function of the algorithm programming iteration for the chosen conductance values. The final programming errors are displayed in the plot. (d) Experimental results of the evaluation of the negated exclusive disjunction (XOR) gate resulting in a 95% classification accuracy. Blue and red symbols correspond to correct and incorrect classification, respectively.

Figure S5c reports representative result of conductance tracking during the running of a program-and-verify algorithm for the 4 conductance values used for the hardware readout experiment. A significant programming error (17.7%) is found only for the lower target conductance (green line). Notwithstanding this result, the XOR classification reaches a 95% accuracy. Indeed, Figure S5d reports the inference experimental results for the XOR task for different input combinations, in terms of current outflowing from the memristor crosspoint architecture, $I_{out}$, compared to a current threshold, $I_{th}$, dashed horizontal line. Only in 6 cases (red dots) over 120 attempts (blue dots), the classification is incorrect and only for the 01-input combination.

Also in this case, no specific optimization has been carried out on the program-and-verify algorithm, which indicates the robustness of the computing system.